\documentclass[%
pra,
reprint,
superscriptaddress,
 amsmath,amssymb,
 aps,
]{revtex4-2}
\usepackage{graphicx}
\usepackage{dcolumn}
\usepackage{bm}
\usepackage{physics, siunitx}
\usepackage{xcolor}
\usepackage{tikz}
\usepackage{pgfplots}
\usepackage{pgfplotstable}
\usepackage{lipsum}
\usetikzlibrary{spy}
\newcommand{\change}[1]{\textcolor{black}{#1}}

\newcommand{\outerprod}[2]{#1 \! \otimes \! #2}

\begin{document}

\title{Transient dynamics of subradiance and superradiance in open optical ensembles}

\author{Elliot Lu}
\affiliation{Department of Physics and Astronomy, Michigan State University, East Lansing, Michigan 48824}
\affiliation{Department of Electrical \& Computer Engineering, Michigan State University, East Lansing, Michigan, 48824}
\author{B. Shanker}%
\affiliation{Department of Electrical \& Computer Engineering, Michigan State University, East Lansing, Michigan, 48824}
\author{Carlo Piermarocchi}
\affiliation{Department of Physics and Astronomy, Michigan State University, East Lansing, Michigan 48824}%


\date{\today}

\begin{abstract}
We introduce a computational Maxwell-Bloch framework for investigating out-of-equilibrium optical emitters in open systems. To do so, we compute the pulse-induced  dynamics of each emitter from fundamental light-matter interactions and  self-consistently calculate their radiative coupling, including phase inhomogeneity from propagation effects. This semiclassical framework is applied to open quantum dots systems with different densities and dipolar coupling. We observe signatures of superradiant behavior, such as directionality and faster decay, as well as subradiant emission. We compare and discuss the computed light emission obtained with our method and a Master equation approach.
Our framework enables quantitative investigations of large optical ensembles in the time domain and could be used to design new systems with enhanced superradiant and subradiant properties.  

\end{abstract}

\maketitle






\section{Introduction}
Superradiance and subradiance in optical systems continue to be under intense experimental and theoretical investigations, both in atomic
and solid-state systems~\cite{scully2009super,cong2016dicke}. 
Theoretical descriptions of these phenomena often rely on effective Hamiltonians, such as the Dicke model, where interaction with one or few cavity modes is assumed and  emitters are homogeneous. For extended systems, Maxwell-Bloch equations can be used where the electric field couples to a continuous local-averaged polarization field ~\cite{hess1996maxwell,jirauschek2019optoelectronic}. However, to understand these collective phenomena, it is essential to consider the role of the emitters' local configuration and their spatial and energy distribution. In fact, the disorder in the local distribution of the emitters can strongly affect the superradiant and subradiant dynamics. The coupling between emitters resulting from the exchange of virtual photons, known as van-der-Waals coupling~\cite{gross1982superradiance},  has been recognized for a long time to be an obstacle to the experimental observation of superradiant behavior~\cite{friedberg1972limited}. Far from being a limitation, subradiance has been recently proposed as a mechanism for photon storage in quantum memories 
~\cite{asenjo2017exponential,orioli2021emergent}. To design new systems that exploit superradiance and subradiance, we need analysis methods that: (1) simulate individual optical emitters and their mutual coupling, (2) include the full spatial dependence of the electromagnetic coupling in three dimensions, and (3) describe the dynamics in the time domain taking into account pulse-induced transients and finite propagation times. 
 


Here, we propose a computational approach to investigate superradiant and subradiant dynamics taking into account local inhomogeneities, propagation effects, and the full spatial dependence of the electromagnetic coupling~\cite{connor_glosser_2018_1246090}. The method relies on an integral formulation of semiclassical  microscopic Maxwell-Bloch equations
~\cite{glosser2017collective}. The numerical solution of such a large number of coupled and time-delayed nonlinear equations in a random medium is challenging~\cite{hoskins2021fast,kraisler2021collective}, and our methods presented in detail in~\cite{glosser2021acceleration} amortize computational cost and accelerate convergence.  In this approach,
the nonlinear dynamics of each emitter and the field generated by the emitters' polarization are self-consistently computed, showing a rich phenomenology of short- and long-lived excitations and synchronized oscillations.   Since the local configuration of emitters can be engineered in solid-state systems, we consider  parameters typical of semiconductor quantum dot systems embedded in a solid matrix, which exhibit strong dipoles and where collective radiative effects have been experimentally observed~\cite{raino2018superfluorescence,miyajima2009superfluorescent, miyajima2011biexcitonic, scheibner2007superradiance}. However, the computational framework is sufficiently general to enable the exploration of other systems, such as optical centers in solids and atomic clouds. For instance, the simulation of individual emitters could be important for interpreting experiments in which the emitters are deterministically placed in the crystal. For example, geometrical arrangements of erbium atoms in silicon nitride systems have been recently studied~\cite{Nandi:20}.
Another group has experimentally demonstrated the concept of “atom-like mirrors,” which trap radiation with deterministically placed artificial atoms acting as resonant mirrors in a waveguide~\cite{mirhosseini2019cavity}. Realistic simulations of superradiant coupling in systems where the emitters have a specific geometry could guide the experimental realization of these ideas in other solid-state systems.

We introduce our Maxwell-Bloch formulation based on an integral representation of the electric field in Sec. \ref{sec:MB}. In section \ref{sec:Res},  we study ensembles containing several hundreds of interacting dots and describe the properties of their emitted field in time and space. Sect. \ref{sec:Comp} provides a comparison of results obtained using the semiclassical and Master equation approaches. The total emission rate is calculated using different initial conditions for the system. In Sect. \ref{sec:Disc}, we discuss our results and offer some conclusions. Appendix \ref{sec:App} contains the derivation of the integral equations in the Rotating Wave Approximation (RWA).
 
\section{Integral Maxwell-Bloch equations}
\label{sec:MB}
We model each emitter as a two-level system interacting with a classical electric field. A collection of density matrices $\rho^i$ represents the quantum state of each dot $i$, and their evolution is governed by the Liouville equations $\hbar \dot{\rho}^i =-i\commutator{\mathcal{H}^i(t)}{\rho^i} - \mathcal{D}\qty[\rho^i]$, where $\mathcal{H}^i$ is the Hamiltonian of the $i$-th dot of energy $\hbar \omega^i_0$ and Rabi energy  $\hbar \chi^i(t) = \vb{d} \cdot \vb{E}(\vb{r}_i, t)$, with $\vb{d}
$ being the dot's transition dipole, and $\vb{E}(\vb{r}_i, t)$ the total electric field at the position of the dot, $\vb{r}_i$. The Lindblad term $\mathcal{D}\qty[\rho^i]$ describes population decay and decoherence, parametrized by $T_1$ and $T_2$. The key idea is solving for the total electric field, $\vb{E}(\vb{r}, t) = \vb{E}_L(\vb{r}, t) + \mathfrak{F}\{ \vb{P}(\vb{r}, t) \}$, self consistently with the Liouville equation. Here $\vb{E}_L(\vb{r}, t)$ is the exciting laser field oscillating at frequency $\omega_L$. The second term, $\mathfrak{F}\{ \vb{P}(\vb{r}, t)\}$, is the radiation electric field due to a polarization distribution $\vb{P}(\vb{r}, t)$ arising from the off-diagonal elements (coherences) of the $\rho^i$ from all the dots. The latter can be written explicitly in integral operator form:
\begin{widetext}
\begin{equation} \label{eq:integral operator}
    \begin{split} 
    \mathfrak{F}\{\vb{P}(\vb{r}, t)\} 
    &\doteq
    - \mu_0 \qty (\partial^2_t \mathrm{I} - c^2 \nabla \nabla ) g \qty (\mathbf{r},t) \star_{st} \mathbf{P} \qty (\mathbf{r},t ) \\
     &= \frac{-1}{4\pi \epsilon} \int_{} ~\Biggl  [
     \left(\mathrm{I} -  \outerprod{\bar{\vb{r}}}{\bar{\vb{r}}} \right) \cdot \frac{\partial_t^2 \vb{P}(\vb{r}', t_R)}{c^2 R} + 
    \left(\mathrm{I} - 3\outerprod{\bar{\vb{r}}}{\bar{\vb{r}}} \right) \cdot \qty(
        \frac{\partial_t   \vb{P}(\vb{r}', t_R)}{c R^2} +
        \frac{\vb{P}(\vb{r}', t_R)}{  R^3}
     ) \Biggr ] d^3 {\vb{r}'} 
     \end{split}
\end{equation}
\end{widetext}
where $\vb{R} = \vb{r} - \vb{r}'$, $t_R = t - R/c$,  $\bar{\vb{r}}=\vb{R}/R$, $g(\vb{r},t) = \delta(t_R)/R$ is the wave equation Green's function, $\star_{st}$ denotes convolution in space and time, and $\vb{P}(\vb{r}, t)=
\sum_{i} \delta(\vb{r}-\vb{r}_i) \: \vb{d} \: \mathfrak{Re}(2 \rho^i_{01}(t))$. 

To effect computational speedup, we transform the equations to a frame co-rotating with the laser frequency $\omega_L$ by writing $\tilde{\rho} = U \rho U^\dagger$, where $U = \text{diag}(1, e^{i \omega_L t})$. The equivalent of Eq.~(\ref{eq:integral operator}) in this rotating frame is
\begin{equation} \label{eq:integral operator rot}
    \tilde{\mathfrak{F}}\{\tilde{\vb{P}}
    \} 
    \doteq
    - \mu_0 \qty (\partial^2_t \mathrm{I} - c^2 \nabla \nabla ) g \qty (\mathbf{r},t) \star_{st} 
    \tilde{\mathbf{P}} 
    \: e^{i\omega_L t} 
\end{equation}
where, after disregarding anti-resonant terms (rotating wave approximation),  $\tilde{\vb{P}}(\vb{r}, t)=
\sum_{i} \delta(\vb{r}-\vb{r}_i) \: \vb{d} \tilde{\rho}^i_{01}(t)$.  We show in Appendix \ref{sec:App} that the radiative Rabi frequency $\chi_{Rad} = \vb{d} \cdot \tilde{\mathfrak{F}}\{\tilde{\vb{P}}\} / \hbar$ can be written as $\chi_{Rad}= -(\Omega + i\gamma)$, where $\gamma$ is a decay term and $\Omega$ is an energy shift. Both terms are proportional to
$\Gamma = d^2 \omega_L^3 / 3\epsilon \hbar \pi c^3$, which corresponds to the  dot spontaneous decay rate for $\omega_L=\omega_0$. This parameter provides a lower bound on decay rates, which can include contributions from other processes (e.g. phonons or Auger processes). In particular, the conditions $1/T_1 > \Gamma$ and $1/T_2 > \Gamma/2$ must hold.
 
The numerical model thus consists of a system of coupled, time-delayed nonlinear Liouville's equations where the right-hand side depends on the derivatives of  
$\tilde{\vb{P}}(t)$ 
up to the second order. Standard methods for solving this system of equations---such as RK4---are unstable. Our approach comprises adapting methods from discretizing time domain integral equations \cite{shanker2009time} and coupling these with a  predictor-corrector scheme to obtain a solution to the coupled Liouville equations~\cite{glaser2009PC,glosser2021acceleration}. At a given timestep $t_n$, the algorithm guesses a value for $\tilde{\rho}^i(t_{n+1})$, then evaluates $\tilde{\mathfrak{F}}\{\tilde{\vb{P}}(t_{n+1})\}$ and $\dot{\tilde{\rho}}^i(t_{n+1})$ to update $\tilde{\rho}^i(t_{n+1})$ until convergence. Repeating this process for each timestep marches the function forward to obtain a solution for all times.

\section{Transient Dynamics} 
\label{sec:Res}
In this section, we illustrate the capabilities of our computational approach  by simulating ensembles of quantum dots embedded in solid media. 
In all simulations, an incident field with the shifted Gaussian waveform:
\begin{equation}
    \vb{E}_{\text{L}}(\vb{r},t)=  E_0  \: e^{-\frac{(\vb{k} \cdot \vb{r}-\omega_L (t-t_0))^2}{2 \sigma^2}} \cos(\vb{k} \cdot \vb{r}-\omega_L t)\: \vu{x}.
\end{equation}
is used to excite an ensemble of dots lying initially in the ground state $(\tilde{\rho}_{00}, \tilde{\rho}_{01})|_{t=0} = (1,0)$.  Here $\omega_L \approx \SI{4.8} \fs^{-1}$, 
$\sigma/\omega_L \approx 
\SI{0.34} {\ps}$
, $\vb{k} = \omega_L / c \: \vu{z}$, and the laser amplitude $E_0$ is adjusted to produce a $\pi$-pulse on each dot in the absence of interactions. First, we consider the case of identical dots with $\omega^i_0=\omega_L$. The resonant pulse is chosen to peak at $t_0 = \SI{5}{\ps}$. The systems of $N$ quantum dots are assumed to be embedded in a  NaCl medium with refractive index $n \approx 1.54$. They are  Gaussian-distributed  with a standard deviation along each dimension of $0.5 \lambda$. This fixed spread implies that the dot density increases with $N$. Each dot has an identical dipole moment $\vb{d}=d\hat{\vb{x}}$, which is varied based on a reference dipole moment of strength 
$d_0/e \approx \SI{2.5}{nm}$.
 For $d = d_0$, decay times $T_1 \approx \SI{8.3}{\ps}$ and $T_2 = 2T_1$ are chosen, and modified for other values of $d$ to satisfy the $d^{-2}$ dependence.

\begin{figure}

    \begin{tikzpicture}
    \begin{semilogyaxis} [axis on top,
        enlargelimits=false,
        ylabel= $\rho_{11}$,
        width=0.45\textwidth,
        height=0.3\textwidth
        ]
        \addplot graphics [xmin=0, xmax=1000, ymin=1e-9, ymax=1e0]{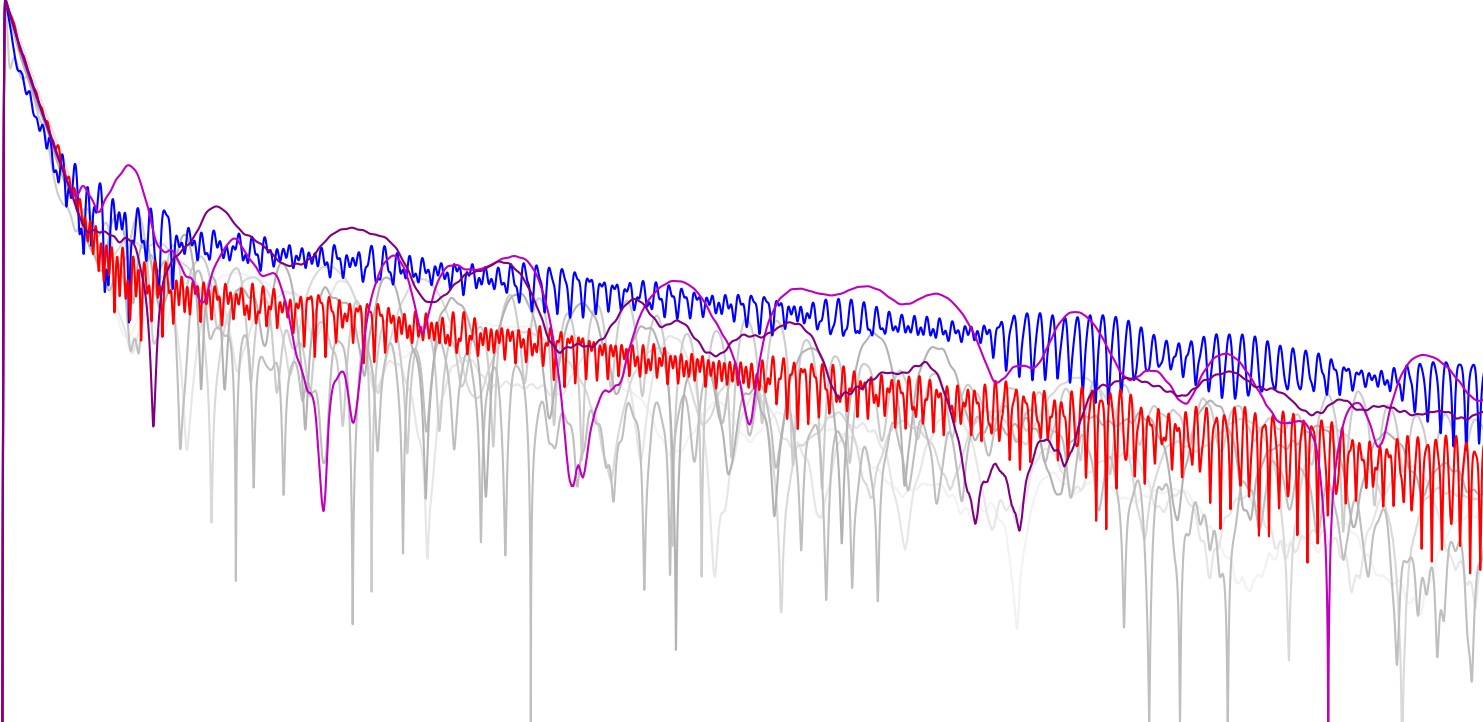};
        \end{semilogyaxis}
        \node at (0.30,0.30) {(\textbf{a})};
    \end{tikzpicture}

    \begin{tikzpicture}[]
    \begin{semilogyaxis}[axis on top,
        enlargelimits=false,
        width=0.45\textwidth,
        height=0.3\textwidth,
        ylabel= $\langle \rho_{11}\rangle$,
        xtick pos=bottom,
        ytick pos=left
        ]
    \addplot graphics [xmin=0, xmax=1000, ymin=1e-7, ymax=1e0]{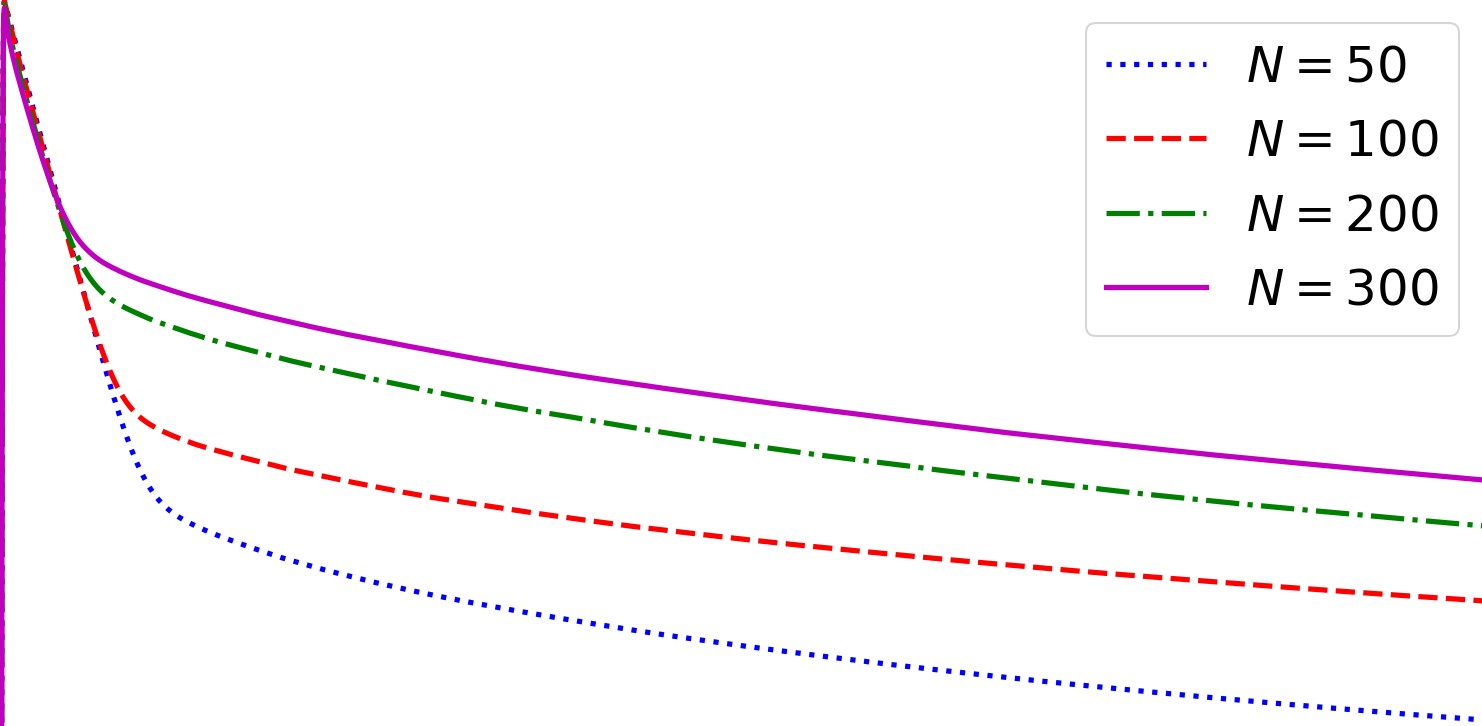};
    \end{semilogyaxis}
    \node at (0.30,0.30) {(\textbf{b})};
    \end{tikzpicture}

    \hspace*{-0.065\linewidth}
    \begin{tikzpicture}[]
    \begin{semilogyaxis}[axis on top,
        enlargelimits=false,
        width=0.45\textwidth,
        height=0.3\textwidth,
        xlabel=Time $(ps)$,
        ylabel= $\langle \rho_{11}\rangle$,
        xtick pos=bottom,
        ytick pos=left
        ]
    \addplot graphics [xmin=0, xmax=75, ymin=1e-3, ymax=1e0]{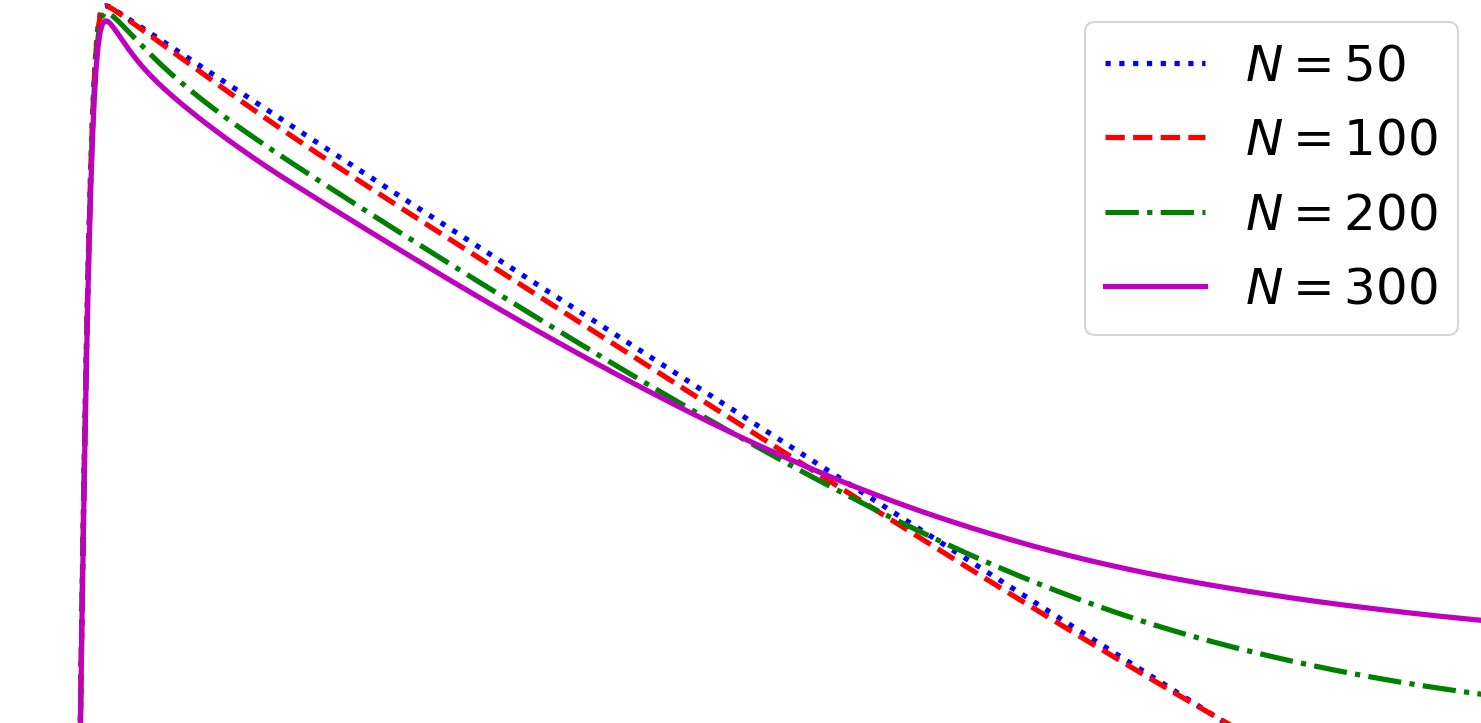};
    \end{semilogyaxis}
    \node at (0.30,0.30) {(\textbf{c})};
    \end{tikzpicture}

    \caption{Time evolution of the population excitation $\rho_{11}$ for a Gaussian distribution of dots with $d/d_0 = 1.0$. (a) Per-dot time evolution for $N=200$. We show here $\rho_{11}^i$,  with $i$ corresponding to ten  randomly chosen dots. (b) Dot-averaged values  $\langle \rho_{11}\rangle$ for different $N$. (c) Immediately post-excitation, a faster decay is observed for larger $N$.
    }
    \label{fig:rho_plt}
\end{figure}
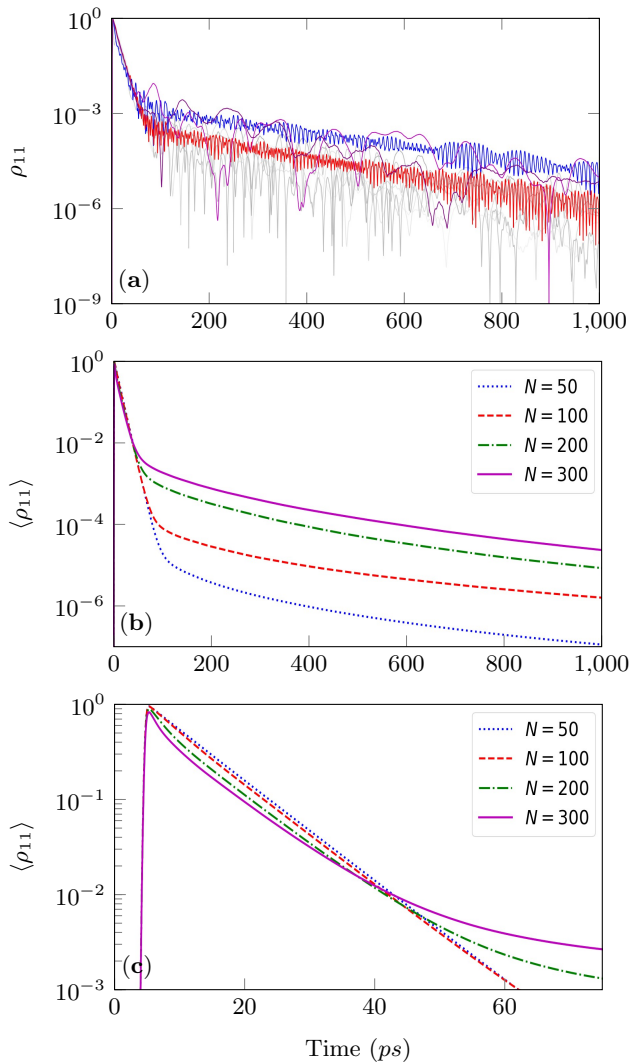

\begin{figure}
    \centering
	 \begin{tikzpicture}
        \begin{axis} [axis on top,
                enlargelimits=false,
                xlabel=$N$,
                ylabel= $d/d_0$,
                ylabel near ticks,
                width=0.39\textwidth,
                height=0.3\textwidth,
                ]
            \addplot graphics [xmin=50, xmax=400, ymin=0.40, ymax=2.0]{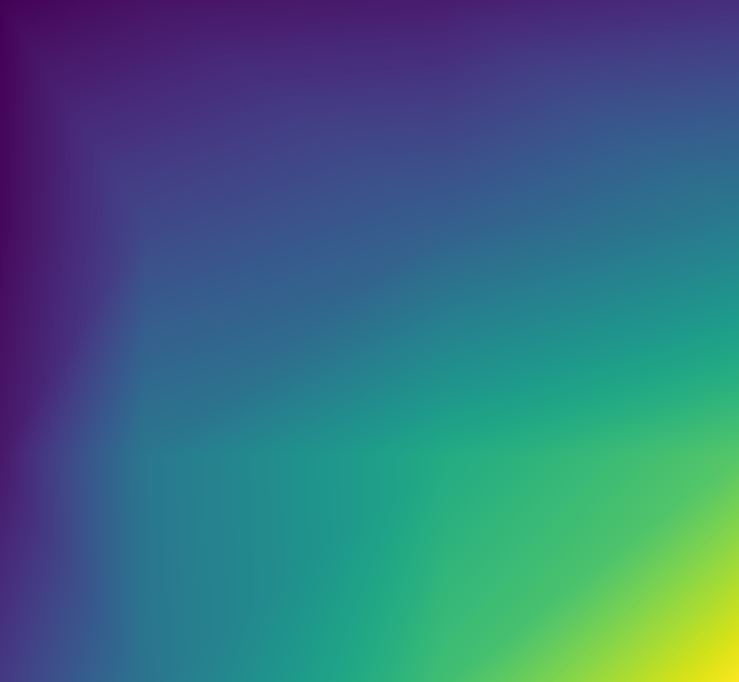};
        \end{axis}
    \end{tikzpicture} 
    \raisebox{0.09\height}{
    \begin{tikzpicture}
        \begin{axis} [axis on top,
                enlargelimits=false,
                xlabel=\change{\scriptsize{$\log \langle \rho_{11} \rangle$}},
                x label style={at={(axis description cs:0.5,-0.00)},anchor=south},
                width=0.10\textwidth,
                height=0.30\textwidth,
                xtick=\empty,
                xticklabels={,,},
                ytick pos=right,
                yticklabel pos=right,
                ]
            \addplot graphics [
            xmin=0, xmax=1, 
            ymin=-7.0, ymax=-3.75]{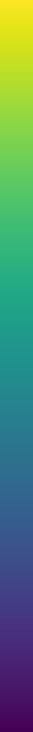};
        \end{axis}
    \end{tikzpicture}
    }

	\caption{ Log-plot of averaged-over-trials population $\langle \rho_{11} \rangle$ at $1$ ns as a function of dipole strength and $N$. }
	\label{fig:Nvsdip_cntr}
\end{figure}

Fig.~\ref{fig:rho_plt}a depicts the time behavior of a set of ten dots chosen from a simulation with $N=200$ dots, portraying a rich phenomenology of oscillations following the initial excitation. The figure shows the excited state population of each dot as a function of time, i.e. $\rho^i_{11}(t)$. These oscillations result from local energy shifts induced by randomly distributed neighboring dots and are dominated by the $1/R^3$ contribution from Eq.~(\ref{eq:integral operator}). After excitation, 
we observe a superradiant behavior, which occurs on the time scale of $T_1$. This can be seen in Fig.~\ref{fig:rho_plt}b-c where we show the population dynamics averaged over all dots, $\langle \rho_{11}\rangle$ for different $N$.  \change{Fig. ~\ref{fig:rho_plt}c displays faster re-emission after pulse excitation in configurations with a greater density of emitters.}
Afterward, the system settles into a subradiant regime where re-emission slows down. This transition is visible in both per-dot (Fig.~\ref{fig:rho_plt}a) and averaged (Fig.~\ref{fig:rho_plt}b) plots. Higher emitter density also leads to more subradiance, resulting in a larger population decaying at long times. Increasing the dipole strength $d$, while increasing the strength of interactions, induces shorter decay times. 
Fig.~\ref{fig:Nvsdip_cntr} summarizes the effects of $N$ and $d$ on the average population at $1$ ns. The subradiant slow decaying states are enhanced by larger $N$ and lower $d$, corresponding to dense systems of weakly interacting dots.

\begin{figure}
    \centering
    \hspace*{-0.015\linewidth}
    \begin{tikzpicture}
    \begin{axis} [axis on top,
            enlargelimits=false,
            xlabel=\change{$\log|\vb{E}|^2$},
            x label style={at={(axis description cs:0.5,5.0)},anchor=south},
            ylabel={\color{white} y ()},
            ytick=\empty,
            xtick pos=upper,
            xticklabel pos=upper,
            width=0.50\textwidth,
            height=0.10\textwidth
            ]
        \addplot graphics [xmin=-25.00, xmax=25.00, ymin=-1.255, ymax=1.255]{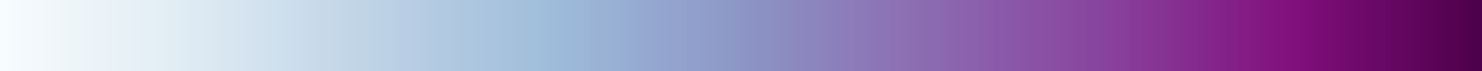};
    \end{axis}
    \end{tikzpicture}
    
    \begin{tikzpicture}
    \begin{axis} [axis on top,
            enlargelimits=false,
            ylabel= y ($\mu$m),
            ylabel near ticks,
            width=0.50\textwidth,
            height=0.40\textwidth
            ]
        \addplot graphics [xmin=-1.255, xmax=1.255, ymin=-1.255, ymax=1.255]{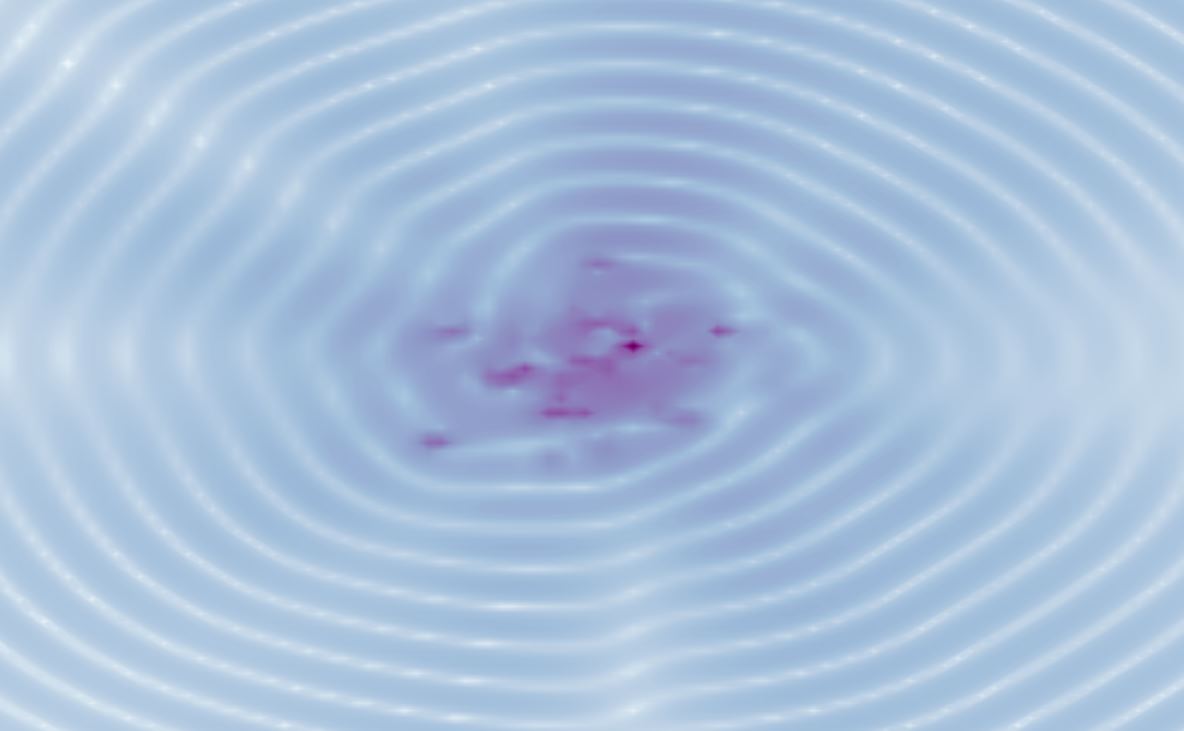};
    \end{axis}
    \end{tikzpicture}
    
    \begin{tikzpicture}
    \begin{axis} [axis on top,
            enlargelimits=false,
            xlabel= x ($\mu$m),
            ylabel= z ($\mu$m),
            ylabel near ticks,
            width=0.50\textwidth,
            height=0.40\textwidth
            ]
        \addplot graphics [xmin=-1.255, xmax=1.255, ymin=-1.255, ymax=1.255]{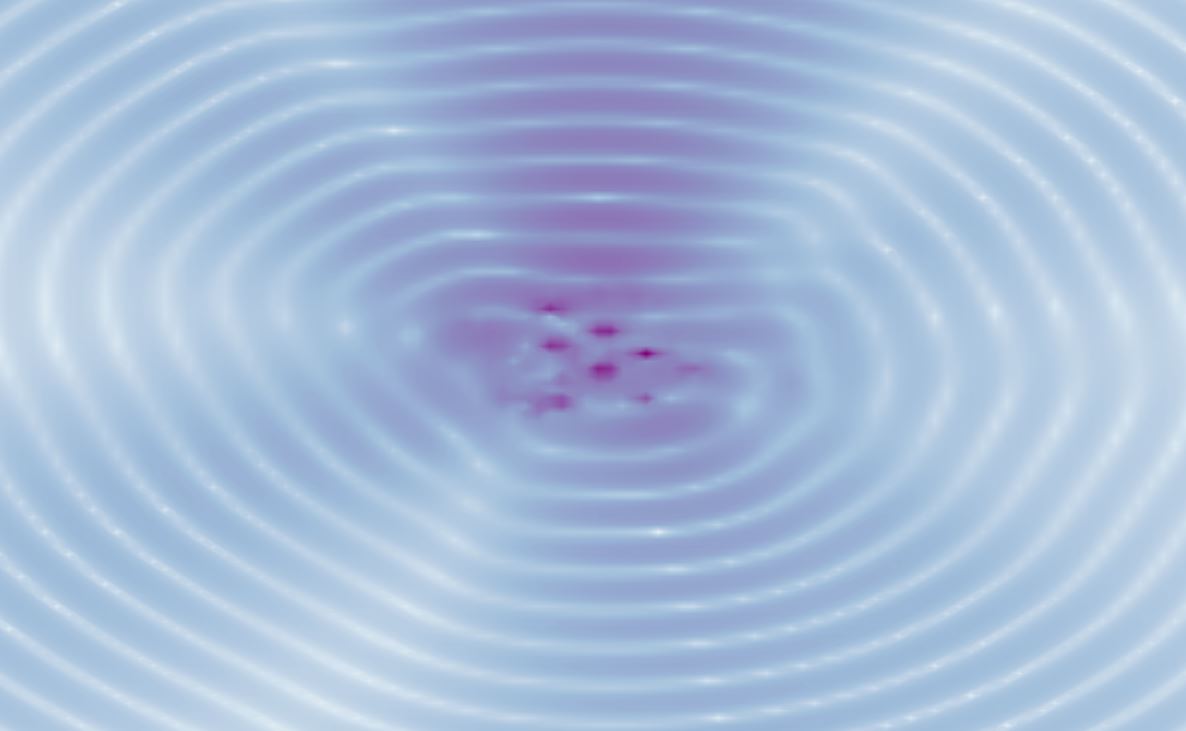};
    \end{axis}
    \end{tikzpicture}

	\caption{Colormaps of logarithmic field intensity (field norm squared) for a configuration with $N=100$ and $d/d_0 = 1.00$ after $20$ ps, on the $x-y$ (top) and $x-z$ (bottom) planes. 
	The spatial oscillations occur with a period about half the wavelength of $\SI{253}{\nm}$. Also note the enhancement along the laser propagation direction in the positive $z$ axis.}
	\label{fig:XY_plt}

\end{figure}

\begin{figure}
    \centering
    
    \hspace*{0.03\linewidth}
    \begin{tikzpicture}
    \begin{axis} [axis on top,
            enlargelimits=false,
            xlabel=\change{$\log|\vb{E}|^2$},
            x label style={at={(axis description cs:0.5,5.0)},anchor=south},
            ylabel={\color{white} y ()},
            ytick=\empty,
            ylabel near ticks,
            xtick pos=upper,
            xticklabel pos=upper,
            xticklabel style={text depth=0pt},
            width=0.45\textwidth,
            height=0.10\textwidth
            ]
        \addplot graphics [xmin=-10.00, xmax=12.00, ymin=-1.255, ymax=1.255]{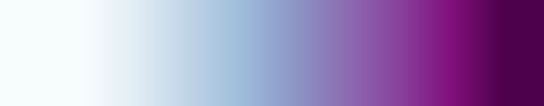};
    \end{axis}
    \end{tikzpicture}
    \begin{tikzpicture}
    \begin{axis} [axis on top,
            enlargelimits=false,
            ylabel=$y$ ($\mu$m),
            ylabel near ticks,
            width=0.45\textwidth,
            height=0.3\textwidth
            ]
            \addplot graphics [xmin=0.0, xmax=200.0, ymin=-2.5, ymax=2.5]{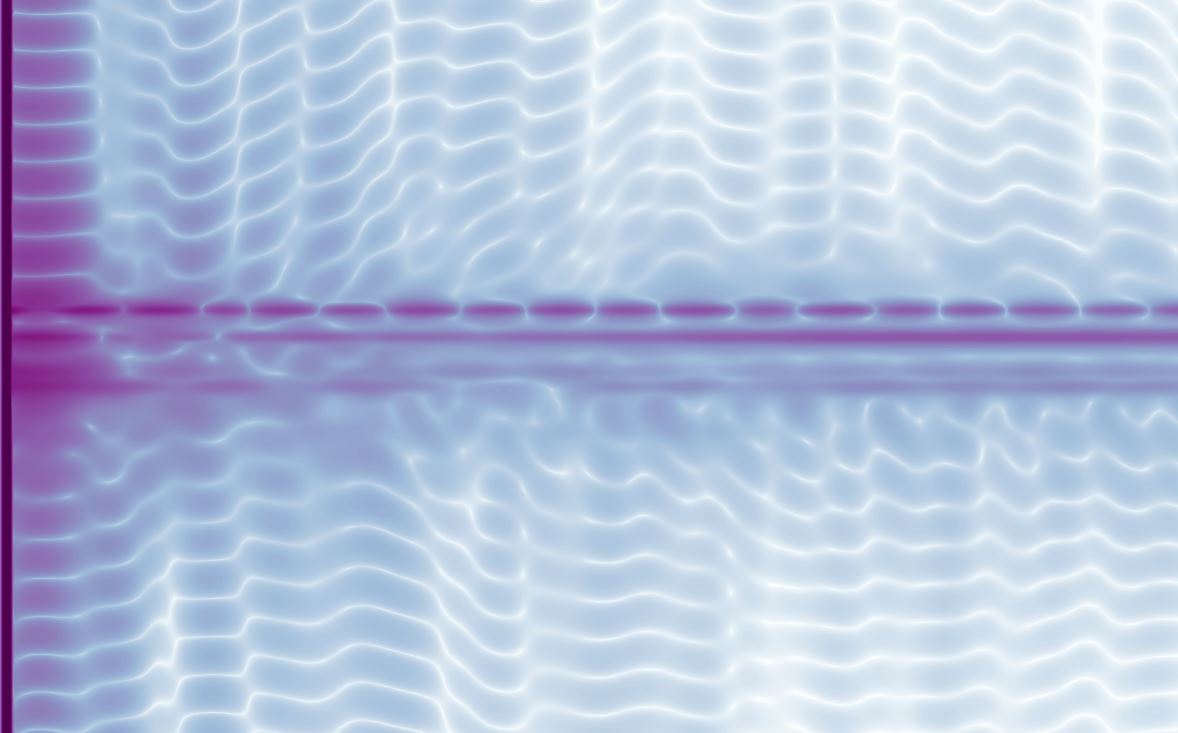};
        \end{axis}
        \node at (6.20,0.30) {(\textbf{a})};
    \end{tikzpicture}
    
    \begin{tikzpicture}
    \begin{axis} [axis on top,
            enlargelimits=false,
            ylabel=$z$ ($\mu$m),
            ylabel near ticks,
            width=0.45\textwidth,
            height=0.3\textwidth
            ]

            \addplot graphics [xmin=0.0, xmax=200.0, ymin=-2.5, ymax=2.5]{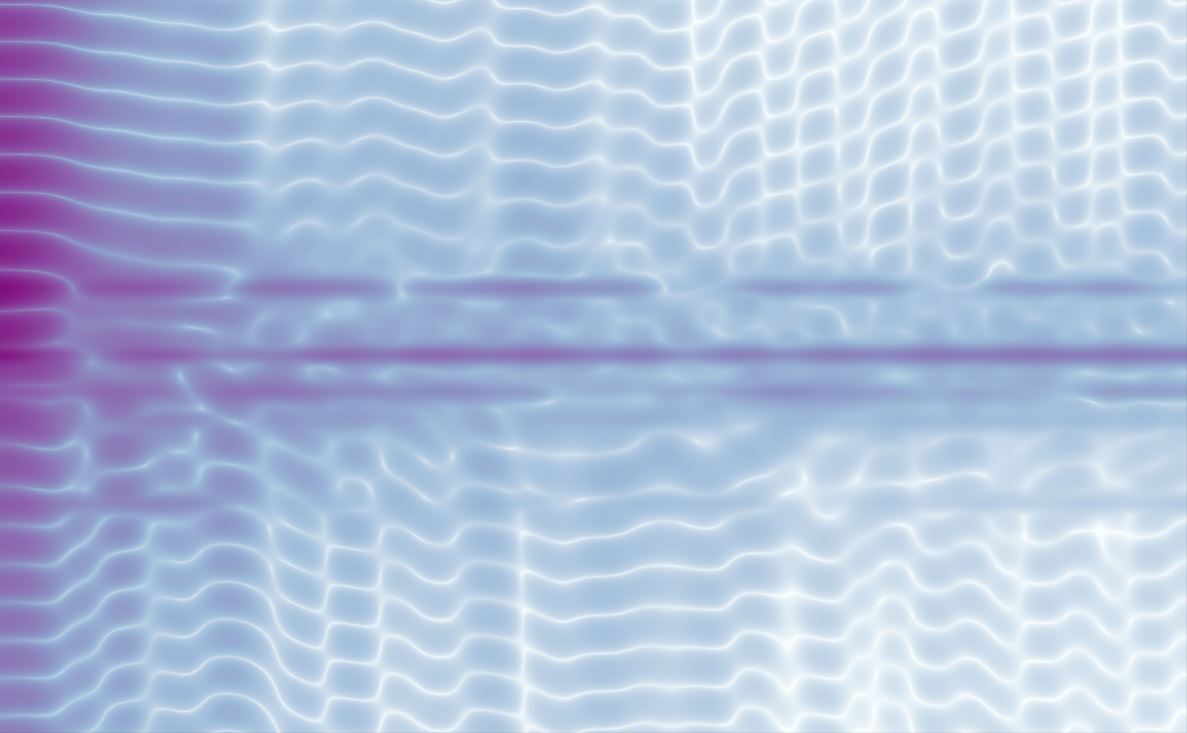};
        \end{axis}
        \node at (6.20,0.30) {(\textbf{b})};
    \end{tikzpicture}
    
    \begin{tikzpicture}
        \begin{axis} [axis on top,
                enlargelimits=false,
                ylabel=$y$ ($\mu$m),
                ylabel near ticks,
                width=0.45\textwidth,
                height=0.3\textwidth
                ]

            \addplot graphics [xmin=0.0, xmax=200.0, ymin=-2.5, ymax=2.5]{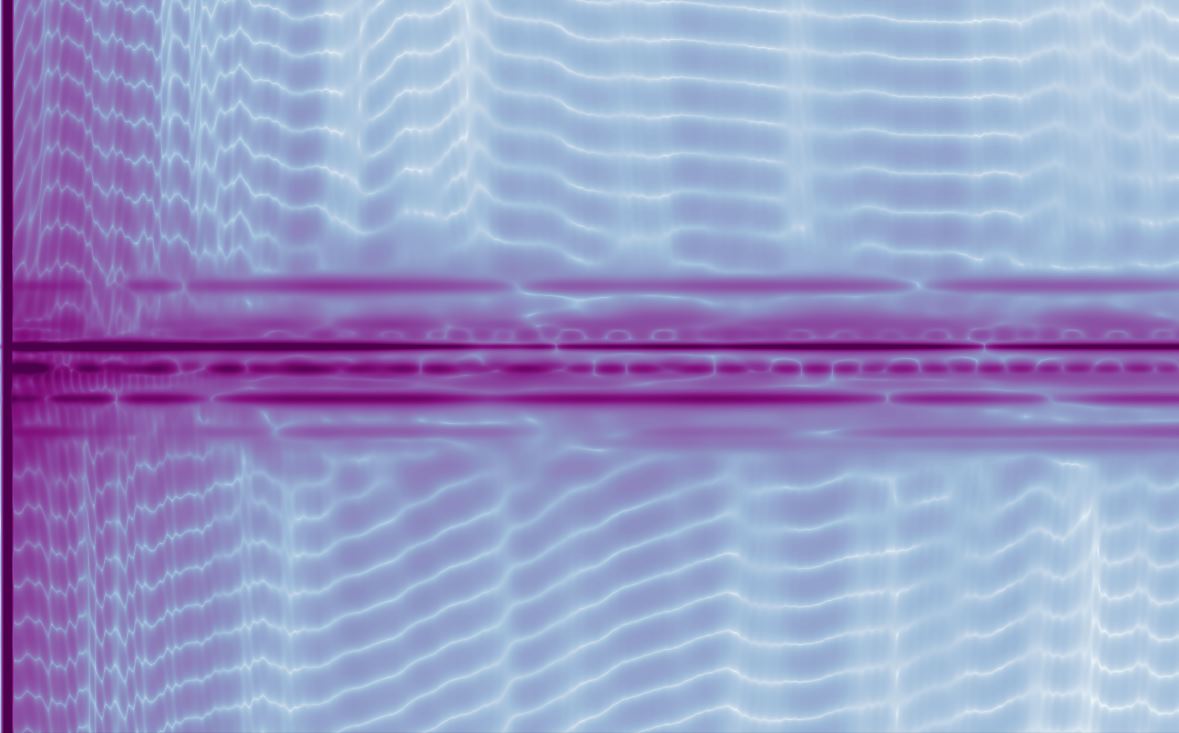};
        \end{axis}
        \node at (6.20,0.30) {(\textbf{c})};
    \end{tikzpicture}
    
    \begin{tikzpicture}
        \begin{axis} [axis on top,
                enlargelimits=false,
                xlabel=Time (ps),
                ylabel=$y$ ($\mu$m),
                ylabel near ticks,
                width=0.45\textwidth,
                height=0.3\textwidth
                ]

            \addplot graphics [xmin=0.0, xmax=200.0, ymin=-2.5, ymax=2.5]{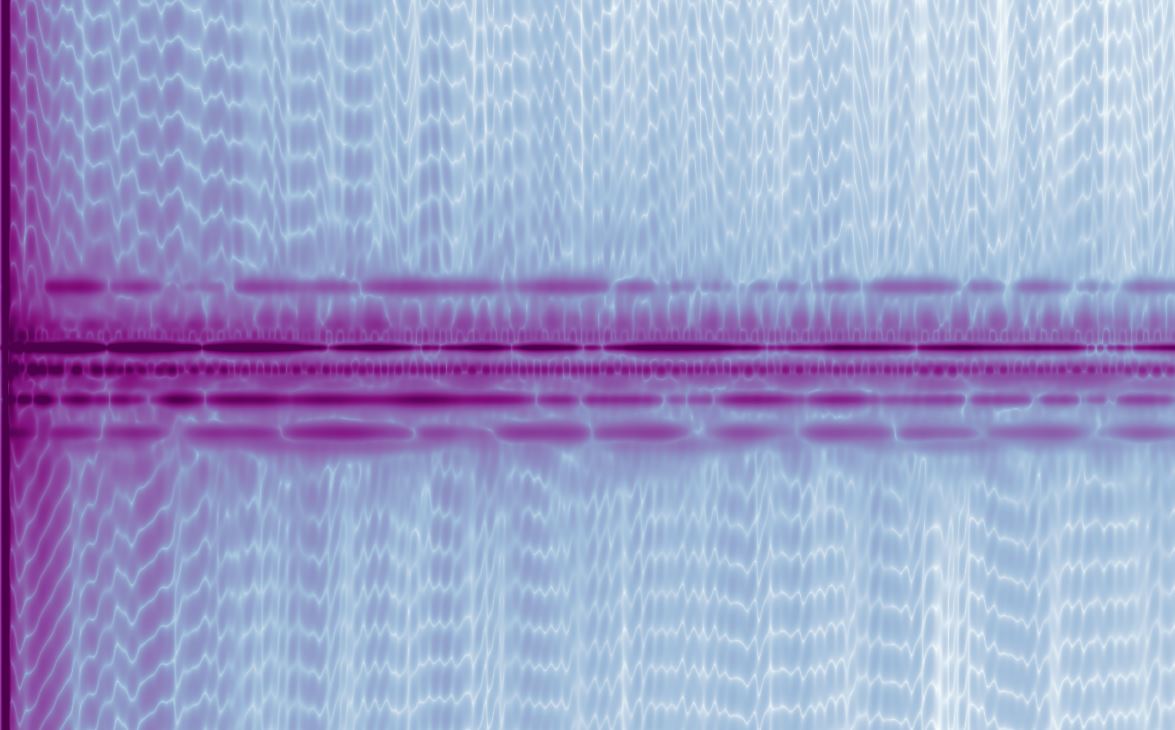};
        \end{axis}
        \node at (6.20,0.30) {(\textbf{d})};
    \end{tikzpicture}

	\caption{Plot of logarithmic radiated field intensities for $N=100$, $d/d_0=1.0$ (a, b), $N=200$, $d/d_0=1.0$ (c), $d/d_0=2.0$ (d) in time and space. Evident are not only spatial but temporal oscillations, becoming enhanced for larger values of the dipole moment. It is also evident that the intensity amplitude increases with the number of emitters. Groups of dots in the cloud ($y \sim 0$) undergo emission synchronization leading to periodic oscillations that become more irregular as the density increase. Finally, (b) displays emission enhancement in the laser propagation direction (cf. Fig 3).}
    \label{fig:TY_plt}
\end{figure}
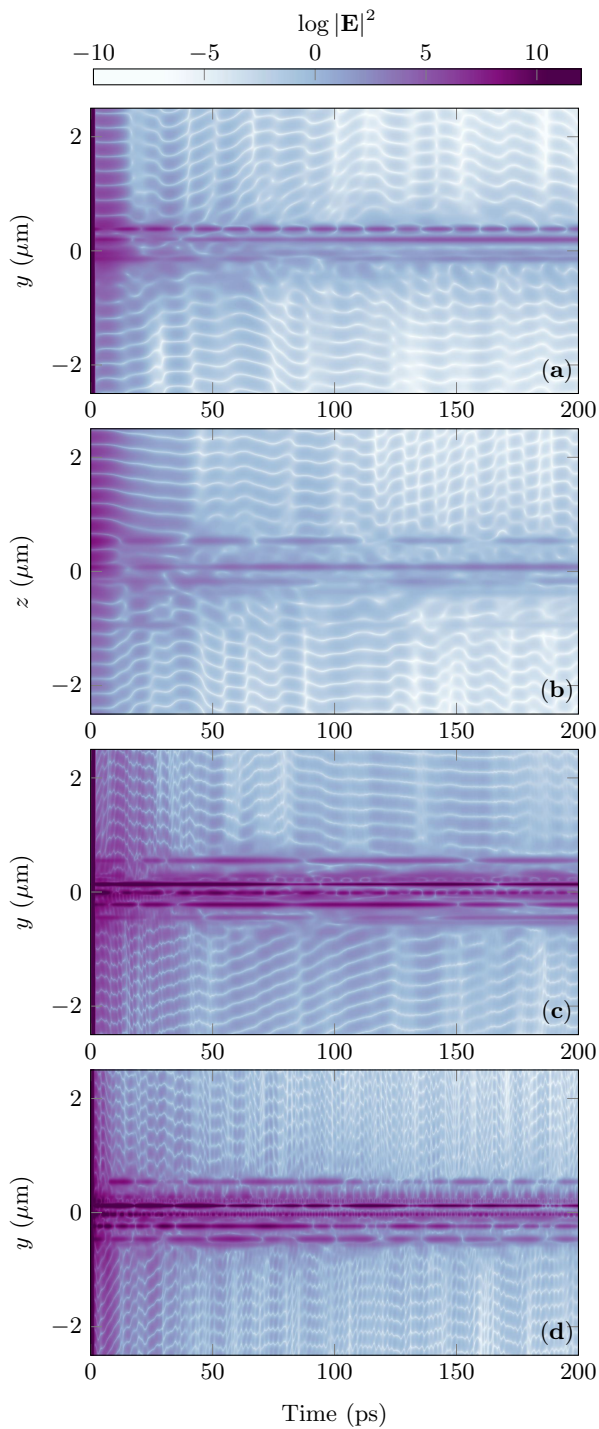

\begin{figure}
    \centering

    \begin{tikzpicture}
        \begin{semilogyaxis} [axis on top,
            enlargelimits=false,
            xlabel=$f$ (ps$^{-1}$),
            ylabel=$|\tilde{I}_1|$,
            width=0.48\textwidth,
            height=0.30\textwidth
            ]
        \addplot graphics [xmin=-1.0, xmax=1.0, ymin=2e-3, ymax=2]{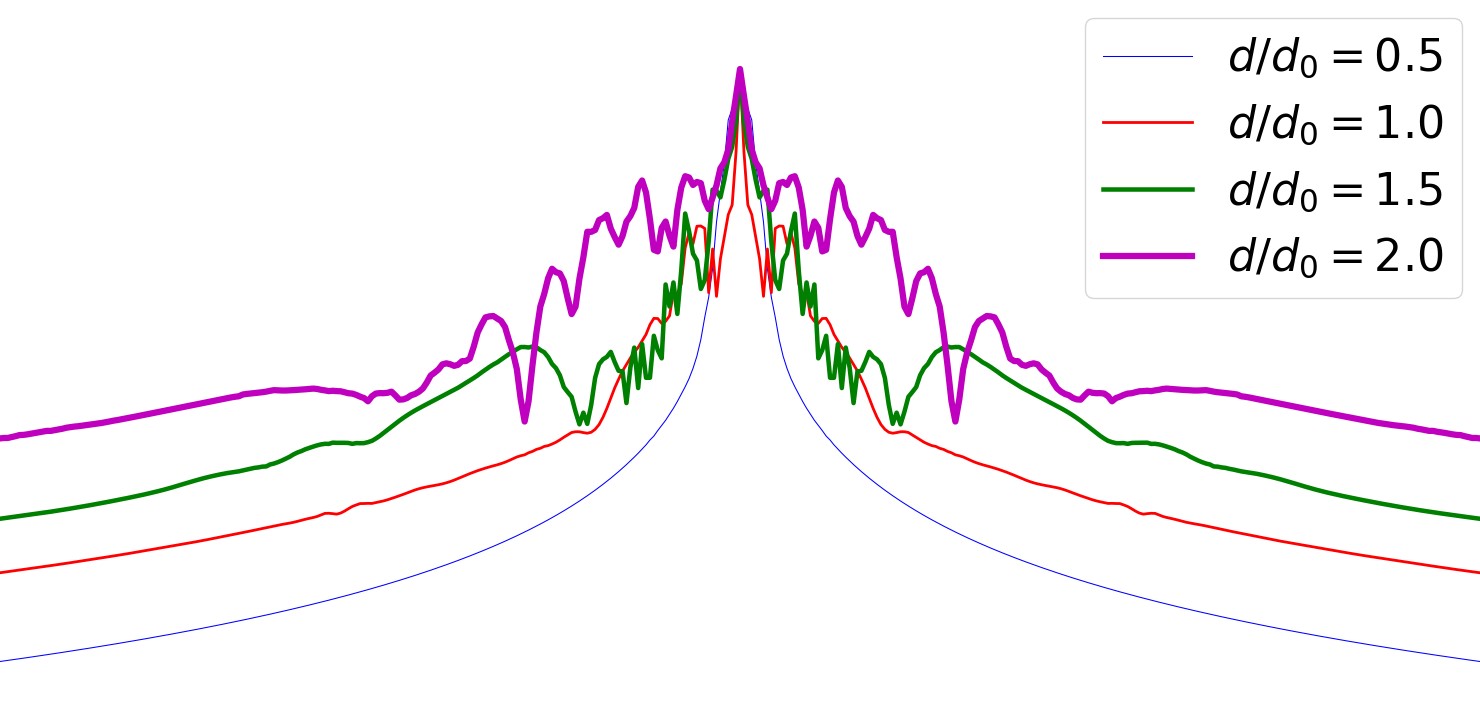};
        \end{semilogyaxis}
    \end{tikzpicture}
    
    \begin{tikzpicture}
        \begin{semilogyaxis} [axis on top,
            enlargelimits=false,
            xlabel=$k$ ($\mu$m$^{-1}$),
            ylabel=$|\tilde{I}_2|$,
            width=0.48\textwidth,
            height=0.30\textwidth
            ]
        \addplot graphics [xmin=-200.0, xmax=210.0, ymin=1e-4, ymax=2]{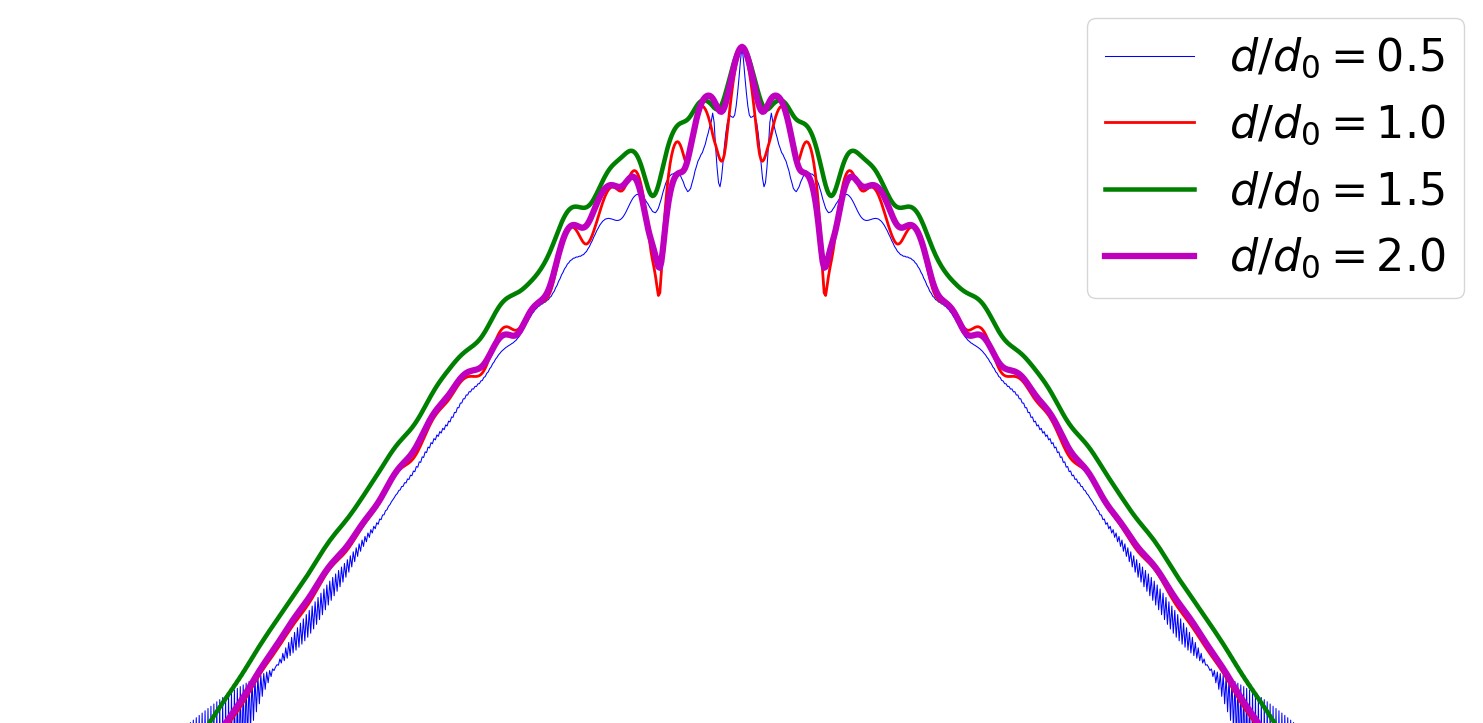};
        \end{semilogyaxis}
    \end{tikzpicture}
    
	\change{\caption{Temporal (top) and spatial (bottom) Fourier plots of field intensity $I(y,t) = |\vb{E}(y,t)|^2$ along the $y$ axis (as in Fig. 4) for $N=200$ and various dipole strengths, normalized by maximum intensity: $\tilde{I}_1(f) = \tilde{I}(k=0, f) / \tilde{I}(0,0)$ (top), $\tilde{I}_2(k) = \tilde{I}(k, f=0) / \tilde{I}(0,0)$ (bottom). Spectral broadening with increasing dipole strength is evident in both cases.}}
    \label{fig:deltavsN_plt}
\end{figure}

Further patterns emerge when examining the radiated field of Eq.~(\ref{eq:integral operator}). Field intensities in the region external to the dot cloud show patterns with a period of half the wavelength (Fig.~\ref{fig:XY_plt}). 
As the excitation pulse propagates along $z$, we also observe a clear enhancement of the emission along the positive $z$ axis, a well-known signature of collective radiative emission~\cite{rehler1971superradiance}.  It is worth noting that the three terms of Eq.~(\ref{eq:integral operator})  are not commensurate. While the near-field interaction terms produce a random pattern near the origin, the far-field $1/R$ term  produces a regular pattern characterized by phase shifts in some directions. 

Time-space plots reveal the synchronization of groups of dots into temporal oscillations in Fig. \ref{fig:TY_plt}, where we plot the intensity along the $y$-axis as a function of time. These oscillations become more pronounced and irregular with increased dipole strength and dot density.
This effect is captured in Fig. 5, which illustrates temporal and spatial Fourier transforms of the field intensity along the $y$-axis for different dipole strengths. These plots display considerable spectral broadening with increasing dipole strength. In time, this broadening is suggested by the emergence of additional peaks corresponding to new oscillation periods in the dot ensemble. Peaks corresponding to characteristic lengths also appear in space, which are nonetheless strongly dependent on the random spatial configuration chosen.

\begin{figure}

    \begin{tikzpicture}
    \begin{semilogyaxis} [axis on top,
        enlargelimits=false,
        width=0.45\textwidth,
        height=0.3\textwidth,
        xlabel=Time $(ps)$,
        ylabel=$\langle \rho_{11}\rangle$,
        ylabel near ticks,
        xtick pos=bottom,
        ytick pos=left
        ]
        \addplot graphics [xmin=0, xmax=1000, ymin=1e-9, ymax=1e0]{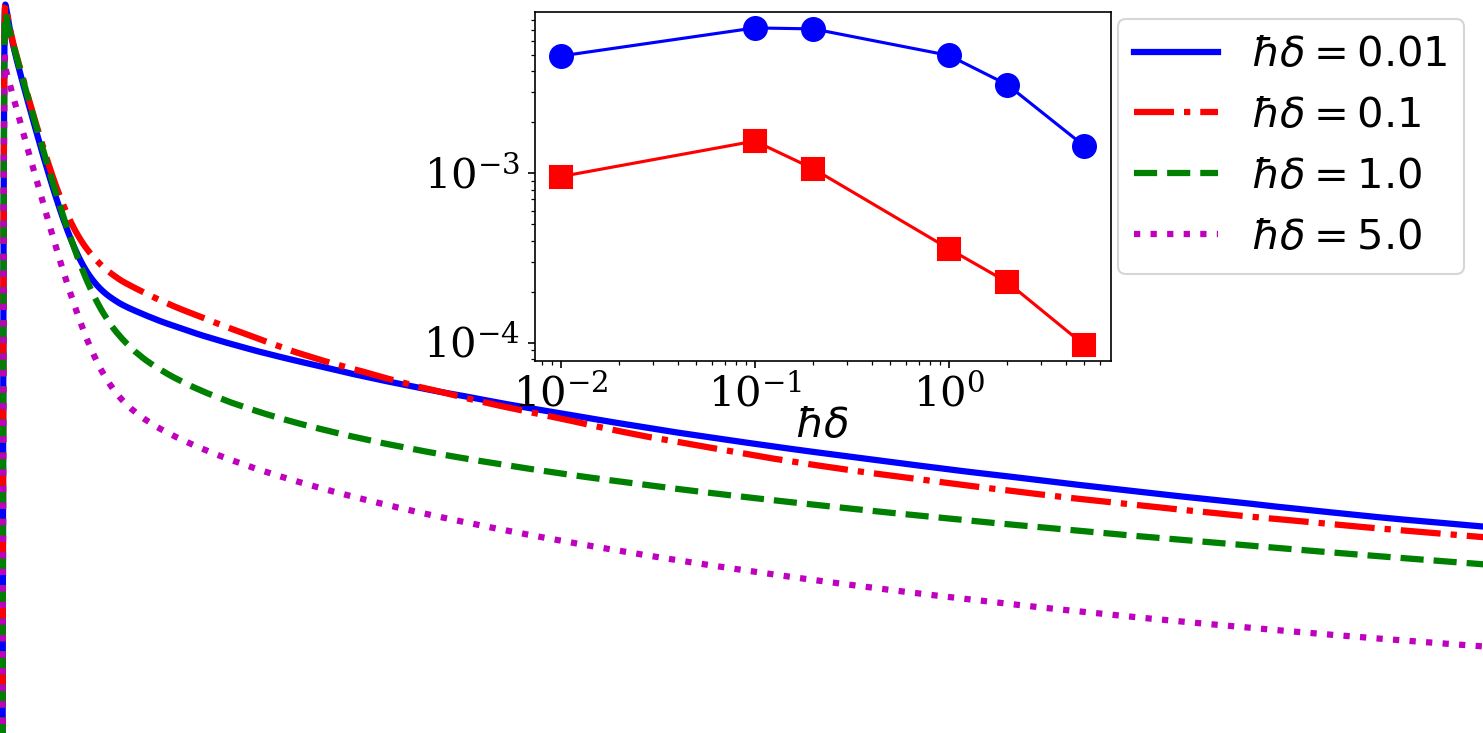};
        \end{semilogyaxis}
    \end{tikzpicture}

    \caption{Time evolution of space-averaged population excitation $\langle \rho_{11} \rangle$ for the same $N=200$ distribution as Figure 1, for different inhomogeneous broadening values $\delta$ (in meV). (Inset) Excitation values \change{$\langle \rho_{11} \rangle$} as a function of $\delta$, at $t = 500$ ps (circles) and $t = 1000$ ps (squares). 
    }
	\label{fig:rho_delta_plt}
\end{figure}

Additionally, we studied the effect of inhomogeneous broadening by considering dots with energy $\hbar \omega_0^i$ that follow a Gaussian distribution of width $\delta$ centered at $\omega_L$.  Increasing $\delta$ affects the dynamics in two ways: (1) the excitation induced by the $\pi$ pulse is less efficient, so the population inversion decreases, and (2) the population of subradiant modes increases due to increased disorder. These two competing effects can be seen in Fig. (\ref{fig:rho_delta_plt}). The average excitation of the dots right after excitation decreases with $\delta$, but at longer times, its dependence on the inhomogeneity is characterized by a peak around $\hbar \delta=0.1$ meV. 

\section{Comparison to Master Equation}
\label{sec:Comp}
Superradiance and subradiance have been described in the literature using two different strategies: the Master equation and the Maxwell-Bloch equation. The former approach traces out the photon degrees of freedom and gives a linear differential equation for the global $2^N \times 2^N$ density matrix \boldsymbol{$\rho$} describing the quantum state of all the two-level systems, which evolves according to
 \begin{equation}
\dot{\boldsymbol{\rho}}(t)=-i [H, \boldsymbol{\rho}] - \sum_{ij} \Gamma_{ij}\left(\{\sigma^+_i \sigma^-_j,\boldsymbol{\rho}\}-2\sigma^-_i\boldsymbol{\rho}\sigma^+_j\right)
\label{eq:ME}
 \end{equation}
 where $ H=\sum_{i\neq j}\Omega_{ij}\sigma^+_i \sigma^-_j$,
 $\sigma^\pm$ are standard Pauli operators and 
 $\{\cdot,\cdot\}$ is the anti-commutator. In Eq.~\ref{eq:ME}, we assumed identical dots and used the RWA (see Appendix) with $\omega_L=\omega_0$. The coefficients $\Omega_{ij}$ and $\Gamma_{ij}$ can be expressed 
 using Eqs.~\ref{eq:G1} and \ref{eq:G2} as $\hat{\boldsymbol{d}_i}\boldsymbol{\Omega}_{ij}\hat{\boldsymbol{d}}_j $ and  $\hat{\boldsymbol{d}_i}\boldsymbol{\Gamma}_{ij}\hat{\boldsymbol{d}}_j$, respectively,  where $\hat{\boldsymbol{d}_i}$ indicate the dipole orientations. To obtain an equation for the global operator $\boldsymbol{\rho}$, one has to rely on a Markov-Born approximation~\cite{agarwal2012quantum} that removes effects due to the finite propagation speed of the electromagnetic field.
 The Maxwell-Bloch approach, on the other hand, can be derived from the Heisenberg equations of motion for the photon and local operators $\sigma^z_i(t)$ and $\sigma^-_i(t)$, and describes memory and propagation effects of the electromagnetic field. The local operators evolve according to delayed differential equations and are coupled in a non-linear way. The quantum Heisenberg equations of motion are formally identical to the semiclassical Maxwell-Bloch equations introduced above if one replaces the  polarization, population, and electric field with the corresponding operators. The theoretical aspects of two approaches have been discussed in the literature and we refer the reader to the review article of Ref.~\cite{gross1982superradiance} for an in-depth analysis.
 
In this section we make a direct numerical comparison of the results obtained with the Maxwell-Bloch predictor-corrector approach introduced in this paper and the Master equation approach. A numerical comparison with $N=100\sim 200$ is computationally impossible due to the exponential scaling of the Master equation approach. The global density matrix $\boldsymbol{\rho}$ contains indeed much more information on the quantum state of the system than what is 
 typically measured in the experiments. We therefore focus on smaller systems with $N=8$ and calculate the transient dynamics of the total radiation emission in the two approaches. Neglecting retardation, which is small for a $N=8$ system, the total photon emission rate can be expressed as
 \begin{eqnarray}
\gamma(t)&=&\sum_i \Gamma_0 \langle \sigma^+_i \sigma^-_i \rangle+ 2 \sum_{i \neq j}\Gamma_{ij}\langle \sigma^+_i \sigma^-_j \rangle  \nonumber \\
&=& \gamma_0(t)+\gamma_I(t)
\label{eq:rate}
 \end{eqnarray}
 where $\Gamma_0=d^2 \omega_0 ^3/3\pi \varepsilon \hbar c^3$. The second term $\gamma_I(t)$ in Eq.~\ref{eq:rate} describes the contribution to the emission from the off-diagonal interactions between the dots and leads to superradiance and subradiance.  In the Master equation approach, the time-dependent expectation values $\langle \sigma^+_i \sigma^-_j \rangle$ are calculated as 
 $\text{Tr} \boldsymbol{\rho}(t)\sigma^+_i \sigma^-_j$ after solving for $\boldsymbol{\rho}(t)$ using Eq.~\ref{eq:ME}. In the quantum  Maxwell-Bloch approach, $\langle \sigma^+_i \sigma^-_j \rangle$ are calculated as $\langle \Psi_0|\sigma^+_i(t) \sigma^-_j(t)| \Psi_0\rangle$, where $|\Psi_0\rangle$ is the state of the system at $t=0$ and the $\sigma^+_i(t)$ 
 are operators in the Heisenberg representation. In the semiclassical approximation, this gives $\langle \Psi_0|\sigma^+_i(t) \sigma^-_i(t)| \Psi_0\rangle \sim \rho^i_{11}(t)$ and $\langle 0|\sigma^+_i(t) \sigma^-_j(t)| 0\rangle \sim {\rho^{i}_{10}}(t) \rho^{j}_{01}(t)$.

\begin{figure}
    \begin{tikzpicture}
    \begin{axis} [axis on top,
        enlargelimits=false,
        width=0.5\textwidth,
        height=0.3\textwidth,
        xlabel=Time $(\text{units of } T_1)$,
        ylabel=Emission rate $\gamma_I(t)$ $(\text{units of } \Gamma_0)$,
        xtick pos=bottom,
        ytick pos=left
        ]
        \addplot graphics [xmin=0, xmax=2, ymin=-0.848, ymax=6.220]{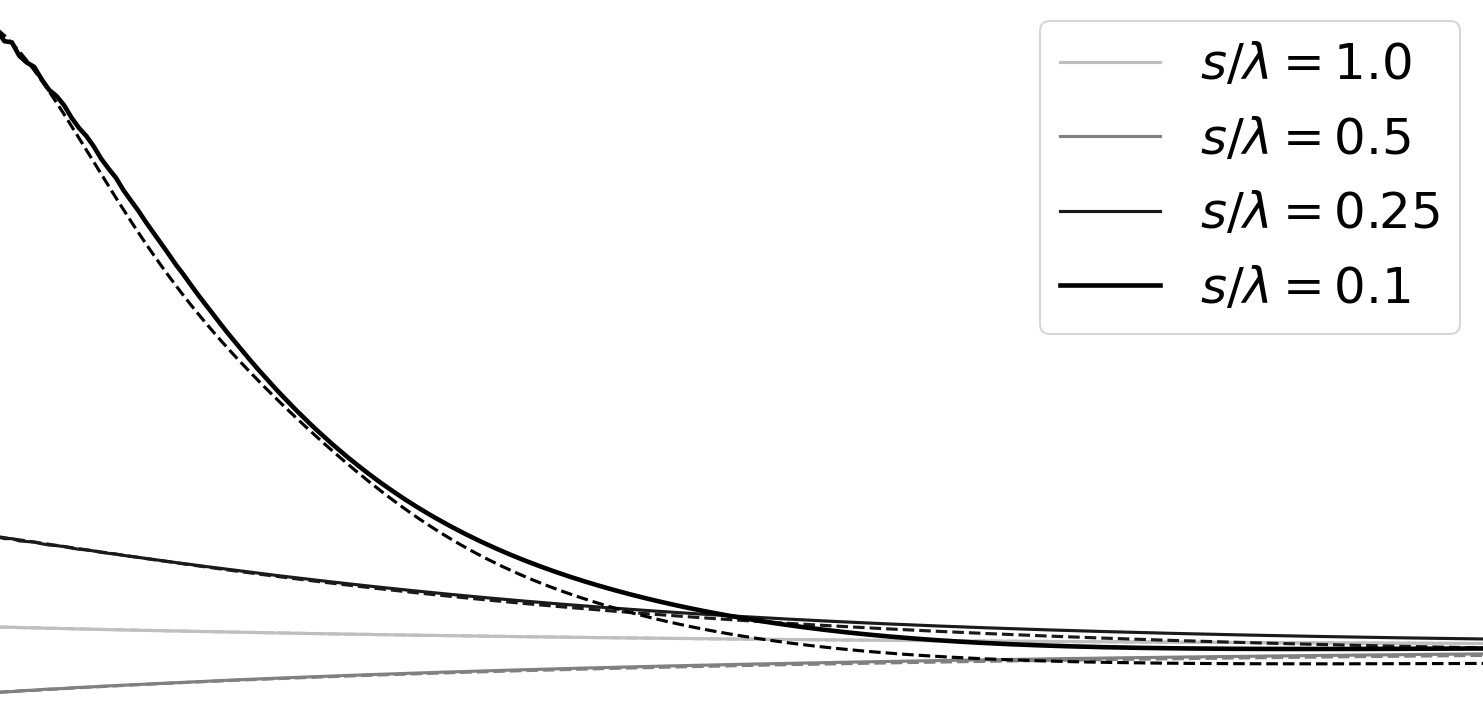};
        \end{axis}
    \end{tikzpicture}
    \caption{Comparison of the off-diagonal term of the emission rate, $\gamma_I(t)$ using the semiclassical Maxwell Bloch (solid) and the Master equation (dashed). Eight quantum dots are initially prepared in the maximum polarization state $(|0\rangle+|1\rangle)/\sqrt{2}$. Curves for different dot separations $s$ compared to $\lambda$ are shown.}
\label{fig:comppi}
\end{figure}

Fig.\ref{fig:comppi} show a comparison of $\gamma_I(t)$ calculated for an initial state of the system in which each dot is in the  $(|0\rangle+|1\rangle)/\sqrt{2}$ state, corresponding to the maximum initial polarization. Eight dots are in a chain along the $y$-axis, equally separated by distances given by $s/\lambda$, in a configuration studied in Ref.~\cite{masson_universality_2022} with the Master equation method. There is a tight agreement between the approaches, and slight deviations are visible only in the  $d/\lambda=0.1$ case. Also evident is the destructive interference that occurs when the dots are separated by a half-wavelength. As expected, for dots separated by $\lambda$ the off-diagonal term becomes negligible in both cases.

A special treatment has to be made if every dot is initially in the $|1\rangle$ state. In this case, the initial polarization of the system is zero and remains zero at all times. 
This initial condition is equivalent to setting the Bloch vector of each dot in the equilibrium ``up" initial position, which is unstable.
The semiclassical approximation would then give no contribution to $\gamma_I(t)$ in Eq.~\ref{eq:rate}. This limitation has been addressed by Haake et al.~\cite{haake1979fluctuations} who showed that semiclassical Maxwell-Bloch equations can describe superradiant pulses if Gaussian-distributed zero-averaging random initial conditions for the polarization are used.  The random initial conditions provide the necessary tipping angle leading to spontaneous emission, and can be seen as the effect of a polarization measurement on the fully inverted state, which gives noise since the population and polarization operators do not commute.

\begin{figure}
    \begin{tikzpicture}
    \begin{axis} [axis on top,
        enlargelimits=false,
        width=0.5\textwidth,
        height=0.3\textwidth,
        xlabel=Time $(\text{units of }T_1)$,
        ylabel=Emission rate $\gamma_I(t)$ $(\text{units of }\Gamma_0)$,
        xtick pos=bottom,
        ytick pos=left
        ]
        \addplot graphics [xmin=0, xmax=5, ymin=-0.754, ymax=3.911]{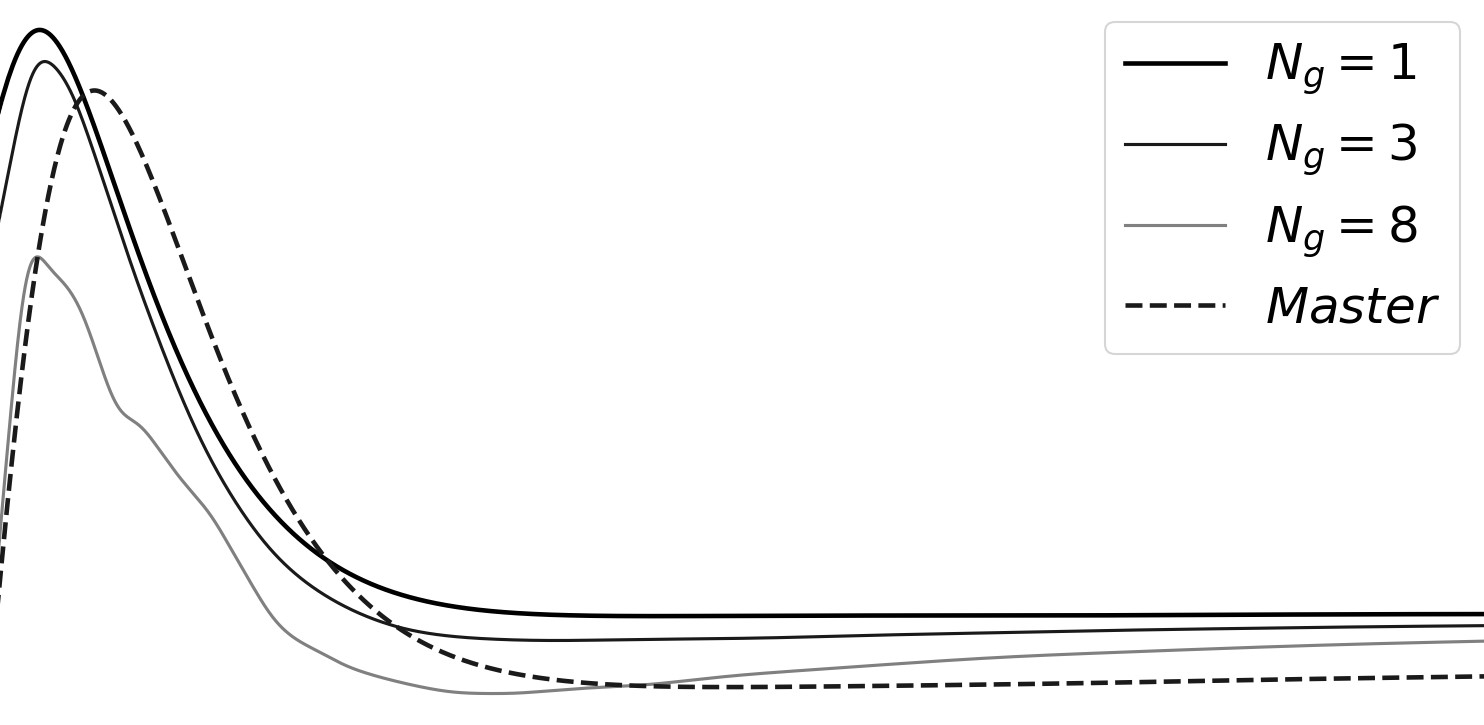};
        \end{axis}
    \end{tikzpicture}
    \caption{Comparison of $\gamma_I(t)$ between the semiclassical Maxwell Bloch and the Master equation. Eight quantum dots equally spaced by 0.1$\lambda$ are considered. Each dot is initially in the maximum population state with the addition of a random polarization (the average has been taken over 500 random initial conditions). Additionally, the dots are divided into $N_g$ groups each receiving a different random polarization phase.}
\label{fig:comppi2}
\end{figure}

Fig.~\ref{fig:comppi2} shows a comparison of $\gamma_I(t)$ for an initial state of the system in which each dot is in the  $|1\rangle$ state, corresponding to the maximum initial population inversion. 
We added a complex zero-averaging polarization following a Gaussian distribution with $\sigma=1/\sqrt{N/N_g}$ according to the method of Ref.~\cite{haake1979fluctuations}. Here $N_g$ is the number of groups into which the dots are divided, each receiving a different random polarization. In general the profile of the emission, consisting of a sharp rise post-excitation followed by decay to a steady low-emission state, resembles that of the Master equation. The methodology of Ref.~\cite{haake1979fluctuations} becomes exact in the limit $N/N_g \gg 1$ and $N_g \gg 1$, therefore the discrepancies observed are due to the fact that we are considering a small system with $N=8$. Small values of $N_g$ lead to spurious non-zero values  for $\gamma_I(t)$ at $t=0$. However, we observe that the averaged initial emission rate correctly tends to zero as the number of groups is increased.

\begin{figure}
    \begin{tikzpicture}
    \begin{axis} [axis on top,
        enlargelimits=false,
        width=0.5\textwidth,
        height=0.3\textwidth,
        xlabel=Time $(\text{units of }T_1)$,
        ylabel=Emission rate $\gamma_I(t)$ $(\text{units of }\Gamma_0)$,
        xtick pos=bottom,
        ytick pos=left
        ]
        \addplot graphics [xmin=0, xmax=5, ymin=-0.4712, ymax=4.712]{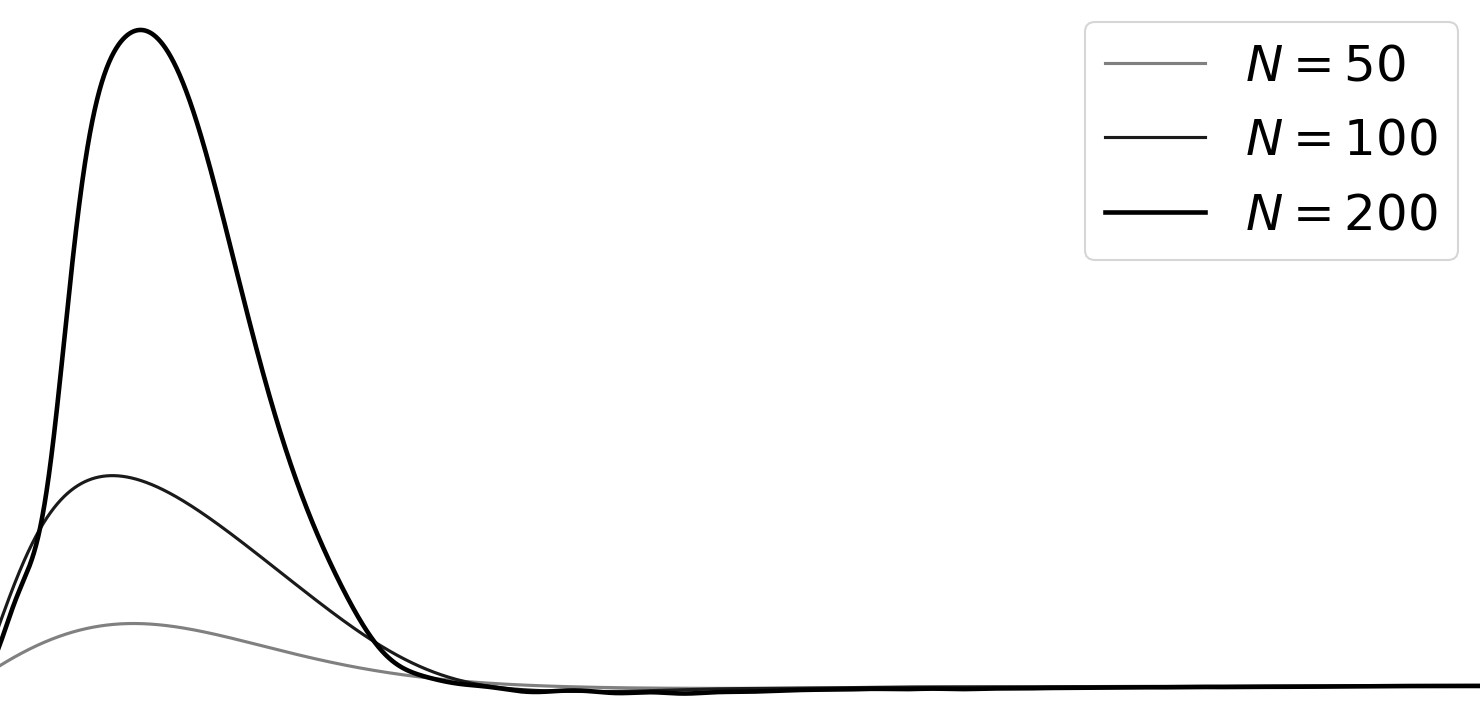};
        \end{axis}
    \end{tikzpicture}
    \caption{The second term of the emission rate, $\gamma_I(t)$, via the semiclassical Maxwell Bloch equations is shown, for the Gaussian distributed dot configurations of Sect. III and all dots starting in the excited state (no pulse, $N_g=1$). The average has been taken over 500 random initial conditions.}
\label{fig:comppi3}
\end{figure}

Using the same zero-averaging of initial conditions, emission rates were finally calculated for the Gaussian distributed dot configurations of Section III. In Fig.~\ref{fig:comppi3}, superradiant emission is evidenced by both the characteristic rise and decay profile, as well as the $N^2$ scaling, of the emission rate. 
(A comparison to the Master equation here is infeasible due to the exponentially scaling computation cost of that approach). We also observed high variability of the emission for different random  configurations in the Gaussian cloud.

\section{Discussion}
\label{sec:Disc}
Our computational approach and software available in~\cite{connor_glosser_2018_1246090} allows us to understand the transient response of ensembles of optical emitters following a laser pulse excitation. We observe superradiant and subradiant emission, and synchronized oscillations. Long-established studies have shown the usefulness of semiclassical models of superradiance~\cite{stroud1972superradiant,rehler1971superradiance}. However, by accounting for the mutual coupling of emitters, our formulation precisely describes local field inhomogeneity. 
We found that superradiance is significantly affected by randomness in electromagnetic coupling. Subradiance leads to a build-up of slowly decaying population inversion, which is maximized at small dipole values and large density, as shown in Fig.~\ref{fig:rho_delta_plt}. Our findings are consistent with recent experiments in Rydberg atoms where the atom-atom electromagnetic coupling led to superradiance decoherence~\cite{suarez2022superradiance}, and the significant deviations observed in quantum dots~\cite{scheibner2007superradiance} from the ideal dependence. When decaying, groups of dots exhibit synchronized oscillations that become irregular for larger dipoles and density, suggesting transitions towards chaotic dynamics~\cite{liu2014dicke}. Energy inhomogeneous broadening contributes to the suppression of superradiance in favor of subradiance. However, we found that the subradiance emission peaks at small levels of energy broadening, as the overall population inversion created by a pulse decreases for large broadening (Fig.~\ref{fig:rho_delta_plt}).

Our comparison of the semiclassical and Master equation approaches has shown similar behavior and good quantitative agreement. An initial condition with perfect population inversion is problematic, as this configuration corresponds to an unstable equilibrium solution for the nonlinear semiclassical approach. However, we have shown how this limitation can be overcome using zero-averaging initial random polarizations. Our coupled nonlinear equations never converge to the unstable equilibrium solution when the initial polarization is different than zero or when random small polarizations are induced by the exciting pulse propagating in the random cloud of dots. 
\change{As shown in the Appendix, the Maxwell Bloch approach uses delayed differential equations to describe interdot-coupling (Eq.\ref{eq:bdelay}), while the Master equation describes the evolution of the density matrix using an ordinary differential equation local in time. This leads to small effects that violate causality, ultimately due to the Markov-Born approximation used in the derivation of the Master equation.  
However, in the 
systems we  studied, these effects were exceedingly small. 
The limitations of a Markovian description of superradiance have been theoretically investigated in the past (see, e.g., Sect. 7 of Ref.\cite{gross1982superradiance}). 
}

To the best of our knowledge, these results are the first large-scale numerical investigation of time-resolved superradiance and subradiance where the dynamics of each emitter is calculated self-consistently. There is extensive literature on superradiance studied with Maxwell-Bloch equations, but always based on macroscopic fields. Due to computational power limitations, our calculations would not have been possible just a few years ago. Also, without our 
Predictor-Corrector-based algorithm, the integration of such an extensive system of non-linear and delayed differential equations would be numerically unstable.  The dependence of slowly decaying states on density and dipole strengths can be understood using general theoretical considerations, e.g., higher disorder leads to more localization and faster emission decreases the population in subradiant states.
However, this is the first time these effects have been explored quantitatively in realistic simulations.


{\it Acknowledgements.} This work was supported by the National Science Foundation  grant OAC1835267 and by Michigan State University through computational resources provided by the Institute for Cyber-Enabled Research.  We thank Dr. S. Masson for providing the software used in Ref.~\cite{masson_universality_2022}. We thank Drs. C. Glosser and M. Maghrebi for enlightening discussions. 

\appendix
\section{\change{Retarded radiated field in the rotating frame}}
\label{sec:App}
Starting from the single dot Hamiltonian
\begin{equation}\mathcal{H}(t) = 
\begin{pmatrix}
	0 & -\hbar \chi(t) \\
	-\hbar \chi^*(t) & \hbar \omega
\end{pmatrix}\end{equation}
and its rotating frame representation
\begin{widetext} 
\begin{equation}\label{eq:rotating_hamiltonian}
\tilde{\mathcal{H}}(t) = U\mathcal{H}U^\dagger - i\hbar U \dot{U}^\dagger= 
\begin{pmatrix}
0 & -\hbar \chi(t) e^{-i\omega_L t} \\
-\hbar \chi^*(t) e^{i\omega_L t} & \hbar (\omega - \omega_L)
\end{pmatrix},\end{equation}
 where $U = $ diag$(1,e^{i\omega_L t})$, we obtain the equations for dot $i$: $\tilde{\rho}^i=U\rho^i U^\dagger$ as 
$\hbar \dot{\tilde{\rho}}^i =-i\commutator{\tilde{\mathcal{H}}^i(t)}{\tilde{\rho}^i} - \mathcal{D}\qty[\tilde{\rho}^i]$, corresponding to \begin{eqnarray}
\label{eq:liouville rot}
\dot{\tilde{\rho}}^i_{00} &=& - i(\tilde{\chi}^{i*} \tilde{\rho}^i_{01}-\tilde{\chi}^i\tilde{\rho}^{i*}_{01}) - (\tilde{\rho}^i_{00}-1)/T_1  \\ 
\dot{\tilde{\rho}}^i_{01} &=& - i(\tilde{\chi}^i (2\tilde{\rho}^i_{00}-1) + \tilde{\rho}^i_{01}(\omega_L - \omega)) -\tilde{\rho}^i_{01}/T_2 
\end{eqnarray}
where $\tilde{\chi}^i = \chi^i \: e^{-i\omega_L t} = \vb{d}^i \cdot (\vb{\tilde{E}}_L + \mathfrak{\tilde{F}}\{\vb{P}\})/\hbar$. 
The Rotating Wave Approximation (RWA) consists in keeping only slowly-varying contributions to $\tilde{\chi}$, which is equivalent to approximating the polarization $\vb{P} =  \vb{d}(\rho_{01}+\rho_{10})\sim \vb{d} \tilde{\rho}_{01} e^{i \omega_L t} \doteq \tilde{\vb{P}} e^{i \omega_L t}$, where $\tilde{\vb{P}}$ identifies a slowly-varying polarization.
The radiated field in the RWA then takes the form 
\begin{equation} \label{eq:integral operator rot}
    \begin{split} 
    \mathfrak{\tilde{F}}\{\tilde{\vb{P}}(\vb{r}, t)\} 
     = \frac{-1}{4\pi \epsilon} \int_{} ~\Biggl  [
     \left(\mathrm{I} -  \outerprod{\bar{\vb{r}}}{\bar{\vb{r}}} \right) \cdot \frac{\partial_t^2 \vb{\tilde{P}}(\vb{r}', t_R) + 2i \omega_L \partial_t \vb{\tilde{P}}(\vb{r}', t_R) - \omega_L^2 \vb{\tilde{P}}(\vb{r}', t_R)}{c^2 R} + \\
    + \left(\mathrm{I} - 3\outerprod{\bar{\vb{r}}}{\bar{\vb{r}}} \right) \cdot \qty(
        \frac{\partial_t \vb{\tilde{P}}(\vb{r}', t_R) + i \omega_L \vb{\tilde{P}}(\vb{r}', t_R)}{c R^2} +
        \frac{\vb{\tilde{P}}(\vb{r}', t_R)}{R^3}
     ) \Biggr ] e^{-i k R} \: d^3 {\vb{r}'} 
     \end{split}
\end{equation}
where $k = \omega_L / c$, $t_R = t - R/c$, and $\vb{\tilde{P}} = \vb{\tilde{P}}(\vb{r}', t_R)$. 
After neglecting terms proportional to $\dot{\tilde{\rho}}_{01}/\omega_L$ and   $\ddot{\tilde{\rho}}_{01}/\omega^2_L$ according to  the RWA, we can write the radiative Rabi energy of dot $i$, $\tilde{\chi}^i_{Rad} = \vb{d}^i \cdot \tilde{\mathfrak{F}}\{\tilde{\vb{P}}\} / \hbar$ as 
\begin{equation}
\tilde{\chi}(t)^i_{Rad} = \sum_{j\neq i}\boldsymbol{\epsilon}_i^*\vb{G}_{ij}\boldsymbol{\epsilon}_j^* \tilde{\rho}^j_{01}(t-R_{ij}/c)
\label{eq:delay}
\end{equation}
where the sum over $j$ extends over all the dots in the systems, the $\boldsymbol{\epsilon}_{i}$ indicate the dot dipole orientations and $\vb{G}_{ij} = \boldsymbol{\Omega}_{ij}-i \boldsymbol{\Gamma}_{ij}$, with
\begin{eqnarray}
  \boldsymbol{\Omega}_{ij}= \frac{3}{4}\Gamma \left( (I - \bar{\vb{r}} \otimes \bar{\vb{r}}) \frac{\cos kR_{ij}}{kR_{ij}} - (I - 3 \bar{\vb{r}} \otimes \bar{\vb{r}}) \left( \frac{\sin kR_{ij}}{(kR_{ij})^2} + \frac{\cos kR_{ij} }{(kR_{ij})^3} \right) \right) \label{eq:G1}\\
  \boldsymbol{\Gamma}_{ij}=\frac{3}{4}\Gamma \left( (I - \bar{\vb{r}} \otimes \bar{\vb{r}}) \frac{\sin kR_{ij}}{kR_{ij}} + (I - 3 \bar{\vb{r}} \otimes \bar{\vb{r}}) \left( \frac{\cos kR_{ij}}{(kR_{ij})^2} - \frac{\sin kR_{ij} }{(kR_{ij})^3} \right) \right)~,
\label{eq:G2}
\end{eqnarray}
\end{widetext}
 where $\vb{R}_{ij} = \vb{r}_i - \vb{r}_j$,  $\bar{\vb{r}}=\vb{R}_{ij}/R_{ij}$, 
and $\Gamma = d^2 \omega_L ^3/3\pi \varepsilon \hbar c^3$. 
\change{Note that Eq.~\ref{eq:delay} in the RWA maintains the delay factor $t-R_{ij}/c$ in calculating the contribution to dot $i$ of the polarization at dot $j$. This preserves causality in light propagation. For instance, consider the case of two dots, $a$ and $b$, in the ground state separated by a large $R\gg \lambda$. Assume that at $t=0$, a small polarization, $\tilde{\rho}^a_{01}$, is excited at dot $a$. In the Maxwell-Bloch equations above (Eqs.~\ref{eq:liouville rot}), the dynamics of the polarization at dot $b$ is described by a delayed differential equation of the form 
\begin{equation}
\dot{\tilde{\rho}}^b_{01}(t) = -i(\Omega_{ba}-i\Gamma_{ba}))\tilde{\rho}^a_{01}(t-R/c)-\tilde{\rho}(t)^b_{01}/T_2.
\label{eq:bdelay}
\end{equation}
The delayed argument on the right-hand side of the equation implies that the polarization at dot $b$  remains exactly zero until the signal from the dot $a$ reaches $b$ after a $R/c$ delay. The Master equation in Eq.~\ref{eq:ME}, on the other hand, is an ordinary linear differential equation, which is local in time. The polarization at $b$, calculated as $Tr_a \sigma_b^+ \boldsymbol{\rho}(t)$, no matter how small, is not identically zero for $t<R/c$.    
}

\bibliography{bibprl}
\end{document}